\documentclass[twocolumn,showpacs,preprintnumbers,amsmath,amssymb]{revtex4}

\usepackage{graphicx}
\usepackage{dcolumn}
\usepackage{bm}
\usepackage{color}

\begin{document}


\title{Single Transverse Spin Asymmetries of Identified Charged Hadrons  \\
in Polarized p+p Collisions at $\sqrt{s}$ = 62.4 GeV}
\newcommand{\bnl}           {$\rm^{1}$}
\newcommand{\krakow}        {$\rm^{2}$}
\newcommand{\newyork}       {$\rm^{3}$}
\newcommand{\nbi}           {$\rm^{4}$}
\newcommand{\texas}         {$\rm^{5}$}
\newcommand{\bergen}        {$\rm^{6}$}
\newcommand{\bucharest}     {$\rm^{7}$}
\newcommand{\kansas}        {$\rm^{8}$}
\newcommand{\oslo}          {$\rm^{9}$}

\affiliation{\bnl}
\affiliation{\newyork}
\affiliation{\krakow}
\affiliation{\nbi}
\affiliation{\texas}
\affiliation{\bergen}
\affiliation{\bucharest}
\affiliation{\kansas}
\affiliation{\oslo}

\collaboration{The BRAHMS Collaboration}\noaffiliation

\author{
I.~Arsene\oslo,
I.~G.~Bearden\nbi,
D.~Beavis\bnl,
S.~Bekele\kansas,
C.~Besliu\bucharest,
B.~Budick\newyork,
H.~B{\o}ggild\nbi,
C.~Chasman\bnl,
H.~H.~Dalsgaard\nbi,
R.~Debbe\bnl,
B.~Fox\bnl,
J.~J.~Gaardh{\o}je\nbi,
K.~Hagel\texas,
A.~Jipa\bucharest,
E.~B.~Johnson\kansas,
R.~Karabowicz\krakow,
N.~Katry\'nska\krakow,
E.~J.~Kim\kansas,
T.~M.~Larsen\nbi,
J.~H.~Lee\bnl,
G.~L{\o}vh{\o}iden\oslo,
Z.~Majka\krakow,
M.~Murray\kansas,
C.~Nygaard\nbi,
J.~Natowitz\texas,
B.~S.~Nielsen\nbi,
D.~Pal\kansas,
A.~Qviler\oslo,
C.~Ristea\nbi,
D.~R{\"o}hrich\bergen,
S.~J.~Sanders\kansas,
P.~Staszel\krakow,
T.~S.~Tveter\oslo,
F.~Videb{\ae}k\bnl,
H.~Yang\bergen,~and
R.~Wada\texas\\ 
  The BRAHMS Collaboration \\ [1ex]
  \bnl~Brookhaven National Laboratory, Upton, New York 11973 \\
  \krakow~Smoluchowski Inst. of Physics, Jagiellonian University, Krakow, Poland\\
  \newyork~New York University, New York 10003 \\
  \nbi~Niels Bohr Institute, Blegdamsvej 17, University of Copenhagen, Copenhagen 2100, Denmark\\
  \texas~Texas A$\&$M University, College Station, Texas, 17843 \\
  \bergen~University of Bergen, Department of Physics, Bergen, Norway\\
  \bucharest~University of Bucharest, Romania\\
  \kansas~University of Kansas, Lawerence, Kansas 66049 \\
  \oslo~University of Oslo, Department of Physics, Oslo, Norway\\
 }


\begin{abstract}
The first measurements of $x_F$-dependent single spin asymmetries of 
identified charged hadrons, $\pi^{\pm}$, $K^{\pm}$, and protons,
from transversely polarized 
proton-proton collisions at  62.4 GeV at RHIC are presented.  
Large asymmetries are seen in the pion and kaon channels.  
The asymmetries in inclusive $\pi^{+}$ production, $A_N(\pi^+)$, increase 
with $x_F$ from 0 to $\sim$0.25 
and $A_N(\pi^{-})$ decrease from 0 to $\sim$$-$0.4.
Observed asymmetries for  $K^-$ unexpectedly
show positive values similar to those for  $K^+$,  increasing with $x_F$, whereas 
proton asymmetries are consistent with zero over the measured kinematic range.
Comparisons of the data with predictions of QCD-based models are presented.
\end{abstract}

\maketitle

The transverse spin dependence of hadron cross-setions in $p^\uparrow +p$ ($\bar{p}^\uparrow+p$)
reactions in the energy regime where perturbative QCD (pQCD) is applicable are expected to be 
negligibly small~\cite{kane} in the lowest-order QCD approximation, whereas experimentally large
left-right asymmetries have been observed~\cite{saroff,adams,bravar} for 
large Feynman-$x$, $x_F = 2p_L/\sqrt{s}$.
Measurements of large asymmetries of inclusive pion production and polarization in
the production of hyperons~\cite{lam} in a wide energy range have motivated various 
theoretical efforts to understand the phenomena. Observed large asymmetries 
are not particularly new phenomena in $p^\uparrow+p$ reactions  since
sizable asymmetries in inclusive pion production had been observed in the 
lower energy regime~\cite{dragoset,antille}, but understanding asymmetries in hadron reactions 
where partonic QCD descriptions are more relevant poses 
a new theoretical challenge.
The main theoretical focus to account for the observed Single Spin Asymmetries (SSAs) in the framework of 
QCD has been on the role of transverse momentum dependent (TMD) partonic effects in the structure 
of the initial transversely polarized nucleon (``Sivers'' mechanism)~\cite{sivers} 
and on the fragmentation process of a polarized quark into hadrons (``Collins'' mechanism)~\cite{collins}. 
Higher twist effects (``twist-3'') arising from quark-gluon correlation effects
beyond the conventional twist-2 distribution have also been considered as a possible origin of 
SSA~\cite{qiu,twist3review}. 
SSA measurements in 
$p^\uparrow+p$ at RHIC energies are of particular interest because 
the next-to-leading-order (NLO) pQCD calculations~\cite{nlo} for 
the unpolarized (spin-averaged) meson cross-sections at forward rapidities successfully 
describe the data~\cite{star,brahms2}.  
At $\sqrt{s}$ $\sim$ 20 GeV where FNAL/E704  observed large SSAs,   
NLO pQCD cross-section calculations~\cite{nlo} significantly under-predict 
the measurements showing increasing discrepancies with increasing $x_F$~\cite{bourrely}.  
The disagreements indicate that there 
may be another mechanism, likely
related to ``soft'' processes, which is significantly responsible for pion
production at this lower energy. 
The two sets of data at $\sqrt{s}$ $\sim$ 20 GeV and at $\sqrt{s}$ $=$ 200 GeV cover 
a similar kinematic  range in $x_F$ and $p_T$ and the measurements show that SSAs for pions 
are energy independent to first approximation.
Since pQCD description of cross-sections at the high-energy region is quite successful, 
while it fails in the low energy domain,  
it might imply that the dominant mechanism responsible for the large SSAs at the two
different energies is a manifestation of two different phenomena.
The newly available measurements from RHIC in the intermediate energy regime 
at $\sqrt{s}$ $=$ 62.4 GeV in $p^\uparrow+p$ can uniquely provide 
an opportunity to clarify the pQCD contribution to SSAs and their energy dependences.   
A simultaneous description of SSAs and the unpolarized cross-sections~\cite{62cross} 
in a wide kinematic range will be a crucial test for the partonic pQCD description.
In particular, flavor dependent SSA measurements allow more complete and stringent 
tests of theoretical models due to flavor dependence in parton distribution functions 
and fragmentation processes. We present here the first measurement of $x_F$-dependent SSAs of 
identified charged hadrons, $\pi^{\pm}$, $K^{\pm}$, and protons,
from transversely polarized proton-proton collisions at  62.4 GeV at RHIC.
\begin{figure}
 \includegraphics[height=0.18\textheight]{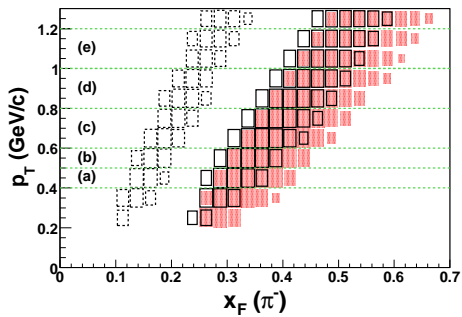}
 \caption{$p_T$ vs. $x_{F}$ for the data used in the SSA analysis. The dotted boxes are 
for the measurements from FS at 6$^\circ$, the filled boxes are from FS at 2.3$^\circ$ 
and the empty boxes with solid line are from 3$^\circ$. 
Data from FS at 2.3$^\circ$ and 3$^\circ$ are used in combination for kaons and protons. 
The size of the boxes represents the relative intensity of the data in logarithmic scale.
The 5 bands marked as (a)-(e) are the $p_T$ ranges used in the Fig.~\ref{fig:pt_dep}. 
\label{fig:acc}}
\end{figure}

The SSA is defined as a ``left-right'' asymmetry of produced particles from 
the hadronic scattering of transversely polarized protons by  unpolarized protons. 
Experimentally the asymmetry can be obtained 
by flipping the spins of polarized protons, and is customarily defined as 
analyzing power $A_N$: 
\begin{equation} 
A_N = \frac{1}{\mathcal P} \frac{(N^+ - {\mathcal L}N^-)}{(N^+ + {\mathcal L}N^-)}, 
\label{eq:An}
\end{equation}
where ${\mathcal P}$ is polarization of the beam, ${\mathcal L}$ is the
spin dependent relative luminosity (${\mathcal L}$ = ${\mathcal
  L_+}$/${\mathcal L_-}$)
and $N^{+(-)}$ is the number of detected particles with beam spin
vector oriented up (down).  
Since both colliding beams are polarized at RHIC, the polarization of ``target'' protons is
averaged over in Eq.~\ref{eq:An}.
The systematic error on the $A_N$ measurements is estimated to be $10\%$ including
uncertainties from the beam polarization, $\delta{\mathcal P}/{\mathcal P}\sim 7.2\%$ 
for the ``Blue'' beam (circulating clockwise) and $9.3\%$ for the ``Yellow'' 
beam (circulating counter-clockwise).   The polarization of the Blue (Yellow) beam is utilized 
for the $A_N$ measurements of particles in positive (negative) $x_F$. The systematic error 
represents mainly scaling uncertainties on the values of $A_N$.  
The average polarization of the beam ${\mathcal P}$ measured by the Hydrogen Jet and pC polarimeters 
is about $50\%$ for the Blue and Yellow beams~\cite{bazil}.

The data presented here were collected by the BRAHMS detector
system~\cite{brahms} with polarized $p+p$ collisions 
from RHIC  with a sampled integrated luminosity of 0.21 pb$^{-1}$ at $\sqrt{s}=62.4$ GeV.
The relative luminosity (${\mathcal L}$) between  the sums of spin-up and spin-down bunches
was measured with a
set of Cherenkov radiators placed symmetrically with respect to the
nominal interaction point~\cite{brahms2}.  The detectors cover the pseudo-rapidity ($\eta$) interval
from $3.26< |\eta| < 5.25$, and are measured from Vernier scans to be sensitive to  
$\sim$$33\%$ of the total inelastic cross-section of 36 mb at 62.4 GeV.  
The uncertainty of determining the relative luminosities is estimated to be 0.3$\%$. 
\begin{figure}
  \includegraphics[height=0.20\textheight]{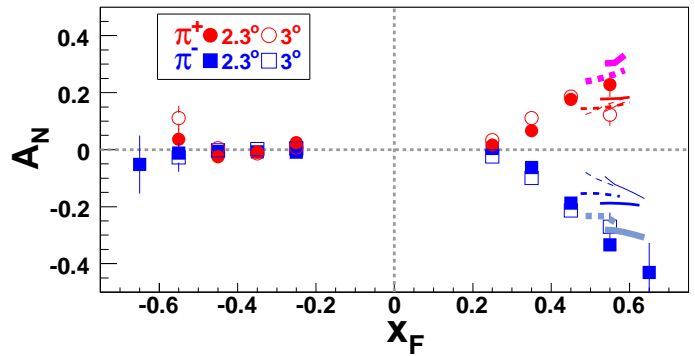}
  \caption{$A_N$  vs. $x_F$ for $\pi^+$ and $\pi^-$. Circle symbols  are for $\pi^+$
and box symbols are for $\pi^-$ measured in FS at 2.3$^\circ$ (solid symbols) and 
3$^\circ$ (open symbols).  The curves are from theoretical calculations. Solid lines are 
to be compared with the data at 2.3$^\circ$ and dotted lines are for 3$^\circ$. 
Thick (solid and dotted) lines are from the initial-state Twist-3 calculations~\cite{feng,feng_c}, 
medium lines are from the final-state Twist-3 calculations~\cite{koike, yuji_c}. 
Predictions from the Sivers function calculations are shown as thin lines~\cite{dalesio,umberto_c}. 
Only statistical errors are shown where larger than symbols.
\label{fig:an_pi_62}}
\end{figure}
\begin{figure*}
  \includegraphics[height=0.2\textheight]{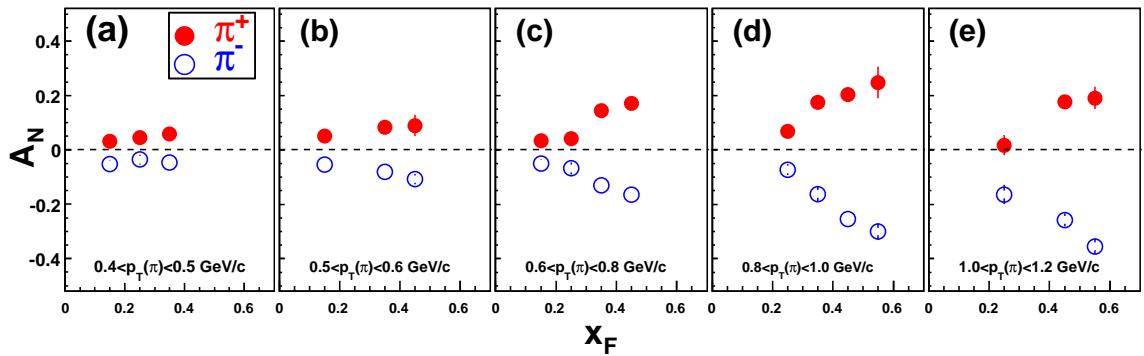}
    \caption{$A_N$  vs. $x_F$  for $\pi^+$ and $\pi^-$
for positive $x_F$ at fixed $p_T$ values:
(a)$0.4<p_T<0.5$, (b)$0.5<p_T<0.6$, (c)$0.6<p_T<0.8$, (d)$0.8<p_T<1.0$,  and (e)$1.0<p_T<1.2$ GeV/$c$ 
as shown in Fig.~\ref{fig:acc}, respectively.  
\label{fig:pt_dep}}
\end{figure*}
The Forward Spectrometer (FS)  measures charged particle
tracks in the forward kinematic region ($\theta$ = 2.3$^\circ$ $-$ 15$^\circ$) 
with good momentum resolution  and particle identification.
The momentum ($p$) resolution of the FS  
is $\delta p/p \approx 0.0016p$ for the half field  
setting where $p$ is in GeV/$c$.
Particle identification was done by utilizing the Ring Image Cherenkov Detector (RICH)~\cite{rich} 
detector which is capable of identifying pions and kaons up to $p \sim 35$
GeV/$c$ and protons above 17 GeV/$c$ 
with an efficiency of $\sim$97\% and a negligible ($\lesssim$0.5\%) probability of misidentification 
in the measured kinematic range ($p$ $<$ 20 GeV/$c$).
The kinematic coverages of the data taken with the FS at 2.3$^\circ$, 3$^\circ$ and 6$^\circ$ 
as a function of $p_T$ and $x_F$ are shown in Fig.~\ref{fig:acc}, where 
the narrow $p_T$-$x_F$ correlated band  at a given setting is due to the small 
aperture of the spectrometer. 
A detailed description of the spectrometer and other experimental details can be found in~\cite{brahms}.

The analyzing power $A_N$ for charged pions, $A_N(\pi^+)$ and 
$A_N(\pi^-)$ at  $\sqrt{s} = 62.4$ GeV as a function of $x_F$ is shown in Fig.~\ref{fig:an_pi_62} for
the two FS angle settings, 2.3$^\circ$ and  3$^\circ$.
At a fixed $x_F$ value, the 3$^\circ$ setting samples higher $p_T$ pions as indicated in Fig.~\ref{fig:acc}.
The mean $p_T$ values $\langle p_T \rangle$ at $x_F$=0.55 are 1.08 and 1.28 GeV/$c$ at 2.3$^\circ$ 
and at 3$^\circ$, respectively~\cite{table}.  The measured $A_N$ values show strong dependence in $x_F$ reaching 
large asymmetries up to $\sim$40\% at $x_F$ $\sim$ 0.6 and no significant asymmetries at $-x_F$.
The decrease of $A_N$ at high-$p_T$ ( $\gtrsim$1 GeV/$c$) and high-$x_F$, especially for $\pi^+$,   
as shown in Fig.~\ref{fig:an_pi_62}
by comparing the two sets of measurements  at 2.3$^\circ$  and at 3$^\circ$ 
might indicate that $A_N$ is in accordance with the expected power-suppressed nature of $A_N$~\cite{feng}.
The asymmetries and their $x_F$-dependence are qualitatively in agreement with the measurements from 
E704 at  $\sqrt{s} = 19.3$ GeV and also most recent $A_N(\pi^0)$ measurements at RHIC  
$\sqrt{s} = 200$ GeV~\cite{adams,star}.  Figure~\ref{fig:an_pi_62} also compares $A_N(\pi)$ 
with a pQCD calculation  in the range of $p_T>1$ GeV/$c$ using ``extended'' twist-3 parton 
distributions~\cite{qiu} including the ``non-derivative'' contributions~\cite{feng,werner,feng_c}. 
In this framework, results of two calculations  from the model 
are compared with the data. One is with only two quark valence densities ($u_v$,$d_v$) in the ansatz,
which is shown in Fig.~\ref{fig:an_pi_62}. The second with additional sea- and anti- quark contributions 
in the model fit slightly increases $A_N(\pi)$ ($\sim$5\%). As the calculations show, the dominant contribution 
to SSA is from valence quarks with contributions from sea- and anti- quarks small enough that 
the current measurements are not able to quantitatively constrain the contribution. 
The calculations, which were done in the same kinematic range as the data, describe the data,
especially  $A_N(\pi^-)$ within the uncertainties.
$A_N(\pi)$ calculated from the ``final-state twist-3''\cite{koike} which uses the twist-3 fragmentation function (FF) 
for the pion  clearly under-predicts $A_N(\pi^{-})$ while is in a reasonable agreement within uncertainties 
for $A_N(\pi^{+})$.
In Fig.~\ref{fig:an_pi_62}, the data are also compared with calculations including Sivers mechanism
which successfully describe the E704 $A_N$ data 
 using valence-like Sivers functions~\cite{dalesio,umberto_c} for $u$ and $d$ 
quarks with opposite sign. The FFs used are from the KKP parameterization~\cite{kkp}, but 
the Kretzer FF~\cite{k} gives similar results.
The calculations underestimate $A_N$, 
which indicates that TMD parton distributions are not sufficient to describe the SSA data at this energy.
As very recent studies~\cite{feng_collins} suggest, Collins mechanism might also be needed to account  
fully for the observed asymmetries. 
All $A_N(\pi)$ calculations compared with the data shows $|A_N(\pi^+)|$ $\sim$ $|A_N(\pi^-)|$ while the data
exhibit $|A_N(\pi^+)|$ $<$ $|A_N(\pi^-)|$ where $p_T\gtrsim 1$ GeV/$c$.
Since there is a strong kinematic correlation between $x_F$ and $p_T$ in the data
as shown in Fig.~\ref{fig:acc}, the rise of $A_N$ in Fig.~\ref{fig:an_pi_62} can
be also driven by $p_T$. 

 Figure~\ref{fig:pt_dep} shows $A_N(\pi^+)$ and $A_N(\pi^-)$ for 5 different
$p_T$ regions from 0.4 to 1.2 GeV/$c$.  As seen in Fig.~\ref{fig:pt_dep}, the $x_F$ dependence of $A_N$ 
at low-$p_T$ ($p_T \lesssim 0.5$ GeV/$c$)
is very small but  increases with $p_T$ in the kinematic region 
at least up to $p_T$$\sim$1 GeV/$c$.    
The $p_T$-dependence of analyzing powers with $x_F$ is qualitatively consistent with the measurements
at $\sqrt{s}$ = 19.3 GeV, where strong $x_F$ dependent SSAs is observed only  above a $p_T$ ``threshold'' 
($\lesssim$ 0.7 GeV/$c$)~\cite{adams}.  It is noted that the trend is also qualitatively in agreement with 
the polarization of the $\Lambda$s produced at the same collision energy,  $\sqrt{s}$ = 62 GeV~\cite{lam}.     
\begin{figure}
  \includegraphics[height=0.18\textheight]{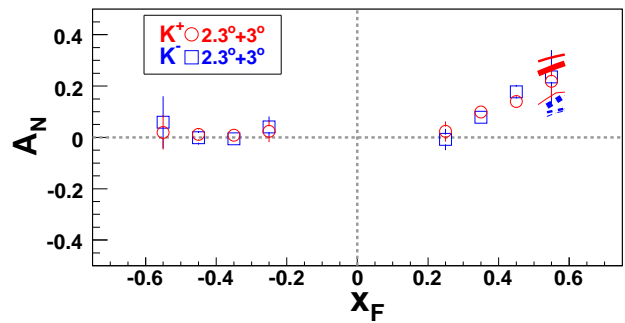}
  \caption{$A_N(K^+)$ and $A_N(K^-)$ vs. $x_F$. Circle symbols are for $K^+$
and box symbols are for $K^-$. 
The solid ($K^+$) and dotted ($K^-$) lines are from the 
initial-state twist-3 calculations with (thick lines) and without (medium lines) 
sea- and anti-quark contribution.
Calculations for the Sivers function are shown as thin lines. 
Errors are statistical only.\label{fig:an_k_62}}
\end{figure}
The SSAs for charged kaons  
as a function of $x_F$ are shown in Fig.~\ref{fig:an_k_62} together with twist-3 and Sivers calculations 
(see the figure caption for details).
The asymmetry for $K^+(u\bar{s})$ is positive as is the $A_N$ of $\pi^+(u\bar{d})$, which is expected 
if the asymmetry is mainly carried by valence quarks, but 
the measured positive SSAs of $K^-(\bar{u}s)$ seem to 
contradict  the n\"{a}ive expectations~\cite{ansel} of valence quark dominance. 
In a valence-like model (no Sivers effect from sea-quarks and/or gluons), 
non-zero positive  $A_N(K^-)$ implies large non-leading FFs ($D_u^{K^{-}}$, $D_d^{K^{-}}$) and insignificant 
contribution from strange quarks. Twist-3 calculations using Kretzer FF also under-predict $A_N(K^-)$ 
due to the small contribution of sea and strange-quark contribution to $A_N$ in the model.
Notably the un-polarized 
cross-section for $K^+$ is an order of magnitude higher than for $K^-$~\cite{62cross}.
The current calculations for kaon asymmetries need an extra or a different mechanism to account 
for positive $A_N(K^-)$ at similar level of $A_N(K^+)$ as shown in Fig.~\ref{fig:an_k_62}. 
If the asymmetries of $K^-$ is mainly driven by pQCD effects, 
the discrepancies between data and calculations are expected 
to be reduced once the Sivers function is better understood for
sea-quarks and also FFs especially the unfavored FFs. 
Likewise
possible non-negligible contributions from the Collins mechanism, as
recently reported~\cite{sidis_collins,hermes2}, may need to be explored further.

SSAs at $x_F < 0$  probe the kinematics of the 
sea (gluon) region of $p^\uparrow$ at small-$x$ and the valence region of $p$, which 
was experimentally measured by the produced particles in the forward hemisphere of $p$ 
in the $p+p^\uparrow$ collisions utilizing the polarization information of the ``target''.  
The measured insignificant $A_N$ for pions and kaons in large $|x_F|$ 
when $x_F < 0$ indicates no significant contribution 
to $A_N$ from processes where $gq$ scattering is enhanced, and the asymmetries are dominated 
by the processes  where large quark PDFs and FFs are expected.
\begin{figure}
  \includegraphics[height=0.16\textheight]{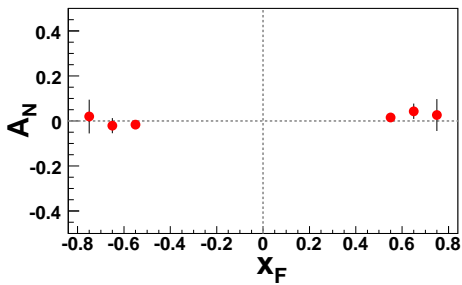}
  \caption{$A_N$  vs. $x_F$ of the proton.
 \label{fig:proton}}
\end{figure}
In Fig.~\ref{fig:proton}, we demonstrate that inclusive protons show no significant asymmetries  
in contrast to pions and kaons in the forward kinematic region.  The insignificant asymmetries
observed are consistent with the measurements at lower energies~\cite{saroff,40gev_proton},
but require more understanding of their production mechanism to theoretically describe the
behavior because a significant fraction of the protons might still be related to the polarized 
beam fragments at this kinematic range~\cite{brahms2}.

In summary, BRAHMS has measured SSAs 
for inclusive identified charged hadron production at forward rapidities in  $p$$^\uparrow$+$p$
at $\sqrt{s}$ = 62.4 GeV.
A twist-3 pQCD model describes the $x_F$ dependence of $A_N(\pi)$ and 
the energy dependence at high-$p_T$ ($p_T>1$ GeV/$c$) where the calculations are applicable, 
but it remains a challenge for pQCD models to consistently describe spin-averaged cross-sections 
at this energy~\cite{bourrely,62cross}.  Measurements of $A_N$ for kaons and protons 
suggest the possible manifestation of non-pQCD phenomena and  call for more
theoretical modeling with improved understanding of the fragmentation processes. 
The energy and flavor dependent asymmetry measurements impose an important  constraint on 
theoretical models describing fundamental mechanisms of transverse 
spin asymmetries  and the Quantum Chromodynamical description of hadronic structure. 

We thank F. Yuan, U. D'Alesio and Y. Koike for providing us with their calculations.   
This work was supported by the office of NP in DoE (USA), 
NSRC (Denmark), RC (Norway), SCSR (Poland), MoR (Romania), and a sponsored 
research grant from Renaissance Technologies Corp.

\end{document}